# Determining adequate consistency levels for aggregation of expert estimates

Vitaliy Tsyganok[1,2,3], Andriy Olenko[4], Pavlo Roik[1] and Oksana Vlasenko[1,2]

[1] *Institute for Information Recording of National Academy of Sciences of Ukraine, M. Shpaka str. 2, Kyiv, 03113, Ukraine*
[2] *Taras Shevchenko National University of Kyiv, Volodymyrs'ka str. 64/13, Kyiv, 01601, Ukraine*
[3] *National Technical University of Ukraine "Igor Sikorsky Kyiv Polytechnic Institute", Beresteysky ave. 37, Kyiv, 03056, Ukraine*
[4] *La Trobe University, Melbourne, Victoria, 3086, Australia*

**Abstract**
To obtain reliable results of expertise, which usually use individual and group expert pairwise comparisons, it is important to summarize (aggregate) expert estimates provided that they are sufficiently consistent. There are several ways to determine the threshold level of consistency sufficient for aggregation of estimates. They can be used for different consistency indices, but none of them relates the threshold value to the requirements for the reliability of the expertise's results. Therefore, a new approach to determining this consistency threshold is required. The proposed approach is based on simulation modeling of expert pairwise comparisons and a targeted search for the most inconsistent among the modeled pairwise comparison matrices. Thus, the search for the least consistent matrix is carried out for a given perturbation of the perfectly consistent matrix. This allows to determine the consistency threshold that corresponds to a given permissible relative deviation of the resulting weight of an alternative from its hypothetical reference value.

## 1. Introduction

To ensure the proper level of quality in decision-making support (DMS), an important aspect is the adequacy of models of weakly structured subject domains on the basis of which decisions are generated. To ensure a sufficient level of adequacy of such models, it is impossible to neglect expert knowledge, which is very significant for any specific application area. In the process of an expert DMS, the adequacy of the models is ensured by the proper reliability of expert estimations obtained during expert expertise. The required level of reliability is achieved by summarizing expert information, which, in general, is redundant. Redundancy of information is inherent in both group assessments, where the assessment of more than one expert characterizes the evaluated object, and individual expert assessments, when a single expert performs pairwise comparisons of objects (alternatives) and the relative weight of each object is determined by more than one pairwise comparison.

Through further aggregation of redundant expert information, new knowledge is generalized and obtained, which forms the basis of a model of a certain subject domain (a complex, weakly structured system). The problem is that generalizing (aggregating, summarizing, averaging) information is advisable only if the set of estimates is sufficiently consistent. Otherwise, in the case of a significant number of contradictions, a situation like the "average body temperature of patients in a hospital" may arise, when this average value is uninformative and does not reflect the real situation, i.e., this value does not convey any information about the system.

Thus, in the process of expert DMS, the task of ensuring a sufficient level of consistency of expert estimates before their aggregation always arises. This directly affects the reliability of the expertise and indirectly the quality of recommendations generated by decision support systems (DSS) and provided to a decision maker (DM). Therefore, a very important aspect of conducting expert assessments is determining the degree of consistency of expert judgments, as well as determining the level of consistency sufficient for aggregating expert estimates.

Therefore, it can be concluded that to obtain reliable results of expertise, it is crucial to take into account the consistency of the expert estimates. The consistency itself does not make much sense without determining a threshold value that serves as an indicator for further aggregation of the estimates. In the case of low consistency of estimates, it is advisable to improve it only to the required level. It is not necessary to achieve full consistency, which, in practice, will lead to an inevitable loss of the effectiveness of the expertise.

✉ vitaliy.tsyganok@gmail.com (V. Tsyganok); A.Olenko@latrobe.edu.au (A. Olenko); paulroyik@gmail.com (P. Roik); vo12062007@gmail.com (O. Vlasenko)

🆔 0000-0002-0821-4877 (V. Tsyganok); 0000-0002-0917-7000 (A. Olenko); 0000-0002-4042-7231 (P. Roik); 0000-0003-4292-6218 (O. Vlasenko)

## 2. The problem

Almost every known method of determining consistency [1, 2] is based on a specific methodology of determining the threshold. Most of them, apart from original approaches such as [3], are based on simulation modeling of expert estimates, for example, the consistency determination by the Analytical hierarchy process method by Thomas Saaty [4]. Existing simulation approaches mostly relate the value of the consistency index to the dimensionality of the pairwise comparison matrix (PCM), but none of them relate this value to the requirements for the reliability of the expertise's results. However, it is intuitively clear that the consistency of the estimates directly affects the reliability ("accuracy") of the expertise results, i.e., affects the closeness of the obtained aggregated estimates to the hypothetical true estimates ("Ground Truth").

## 3. The proposed approach

The proposed approach is based on the postulate of the existence of a "fundamental truth" (Ground Truth), which, in fact, makes any expertise possible. This postulate suggests the existence of reference weights of alternatives – their true estimates, which should be determined in the process of the expert assessment. Thus, it is proposed to set arbitrary weights of alternatives (we assume that they are a priori known).

Next, expert estimates are modeled as pairwise comparisons of a set of alternatives according to a certain criterion. This simulation is carried out as follows:

Given arbitrary weights $w_i, i \in \{1..n\}$, where $n \in \mathbb{N}$ – is the number of alternatives, a fully (perfectly) consistent PCM $M = \{m_{ij}\}$, $i,j \in \{1..n\}$, where $m_{ij} = \frac{w_i}{w_j}$ is constructed. It is easy to see that the matrix $M$ is multiplicative, inverse symmetric: $m_{ij} = 1/m_{ji}$.

Then a noisy version the matrix $M$, $M^\delta$ is created by perturbing each element of the matrix $M$ with some perturbation $\delta$, e.g, $m_{ij} = m_{ij} \pm m_{ij} \cdot \delta$. In essence, $\delta$ acts as a relative estimation error inherent in the expert's assessment when performing pairwise comparisons.

The simulated PCMs can be aggregated using various methods [5, 6] and the result of such aggregation (as well as the result of the expertise) is a vector of priorities of alternatives – their relative estimates. In addition, such generated PCMs can serve as input data for methods for determining consistency. In general, the consistency of the PCM can be assessed using any of the known consistency indices for which the consistency threshold would be determined.

In this simulation experiment, we use the consistency index proposed in [7, 8], which fully meets the requirements for organizing expert feedback, necessary in the case of increasing consistency. This index determines the consistency of the set of estimates generated by the PCM in the approach proposed in [9]. For each alternative to be evaluated, a set ("spectrum") of estimates is built as follows. This set of estimates includes components, each of which is obtained from a separate spanning tree of the graph corresponding to the PCM. Thus, the corresponding consistency index is calculated for the set of scores of each alternative.

In this case, an expression that takes into account the sum of the distances between expert estimates for all possible pairs of estimates is considered as a consistency index:

$$I_a = 1 - \frac{\sum_{i \neq j} f(|x_i - x_j|)}{M}, \quad (1)$$

where $x_i$ is $i$-th expert estimate in the distribution of estimates, $M$ is the value for the most inconsistent case, i.e. the value that maximizes the numerator of the fraction in the expression.

The resulting consistency index $I$ for the PCM is the minimum value of the consistency index among the values given by formula (1) for all alternatives' estimates:

$$I = \min_{a \in A} I_a, \quad (2)$$

where $A$ is the set of alternatives to be evaluated.

The approach to determining the consistency threshold, as well as to evaluating methods of aggregating expert estimates [5, 6], uses a targeted search performed by the Genetic Algorithm [10]. Among the modeled PCMs, it finds those that after aggregation cause the greatest deviation $\Delta$ weights of the alternatives from the reference values. They also represent the most inconsistent case of pairwise comparisons (the lowest consistency index $I$).

A similar approach was used to determine the effectiveness of pairwise comparison aggregation methods [5, 6]. Based on the results of this preliminary study, it is proposed to use the obtained dependence $\Delta(\delta)$ for the Combinatorial method [11, 6], which proved to be the most effective among other aggregation methods.

Thus, as a method for aggregating expert multiplicative pairwise comparisons, we consider the Combinatorial Method [11], the essence of which is to decompose the graph corresponding to the PCM into spanning trees and derive the priority vectors for each such tree, followed by elementwise aggregation of these vectors.

## 4. Input experimental data

The input data, the proposed PCMs, were obtained at the values of the reference (etalon) weights: $w = \{1, \sqrt{3} = 1.732051, \sqrt{3}\sqrt{3} = 3, 3\sqrt{3} = 5.196152, 3\sqrt{3}\sqrt{3} = 9\}$, as an imitation of arbitrary values set on an expert assessment scale limited in the range $[1, 9]$. In the study, see [5, 6], it was experimentally confirmed that the obtained dependencies $\Delta(\delta)$ for each individual method of aggregation only to some extent (not significantly) depend on the specific values of the input reference weights of the alternatives $w_i, i = \overline{1..n}$. That is, setting specific values of input weights and their ordering does not lead to a loss of generality of conclusions in this study. The main requirement for the input reference weights, in addition to the fact that they are real positive numbers in the range $[1, 9]$, is the requirement that they are of the same order of magnitude [12]. This requirement is common to ensure the reliability of the results of expertise based on pairwise comparisons and is one of the conditions for the correct application of the Analytical hierarchy process method [13].

Thus, one of the main results of the simulation study in [5, 6] (with these given reference weights), namely, the experimentally determined dependence of the maximal possible relative deviation of the weights of alternatives $\Delta$ on the relative assessment error of expert $\delta$, is shown in Figure 1.

Based on the same input data that were previously used in the experiment to obtain the dependence $\Delta(\delta)$, it is proposed to used the modified program module to conduct a similar experimental study to obtain the dependence $I(\delta)$.

The available toolkit for the experiment included the implementation of evolutionary methods for the targeted search for optimal solutions. The toolkit was used to find the maximum possible relative deviation $\Delta$ for each $\delta$. This search was carried out using the Genetic Algorithm (GA) [10]. We used the method implemented in a software module, that uses single-point crossover and mutation. The fitness function was defined as the computation of the relative deviation $\Delta$ (the sum of the absolute values of the deviations of the weights of alternatives calculated as a result of the aggregation of the PCMs from the specified reference values).

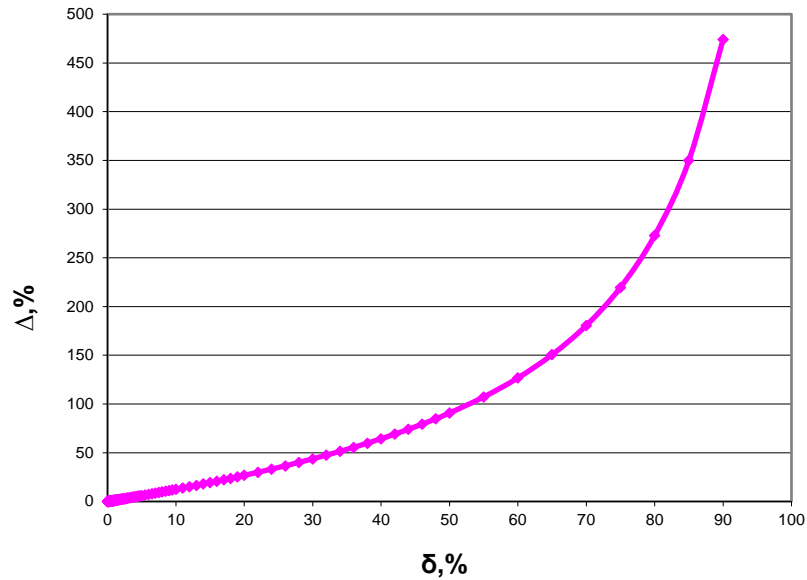

**Figure 1:** Results of the experimental study of the Combinatorial method of pairwise comparison aggregation

Two individuals are randomly selected from the population for crossbreeding. As a result of reproduction, an offspring ("child") appears, which replaces the least adapted individual (with worse fitness).

The module receives the vector of specified input weights from the input file. The computation results $\Delta$ are saved in another file for further analysis. The module control interface is built as a dialog box (see Fig. 2) for entering the necessary data and experimental parameters.

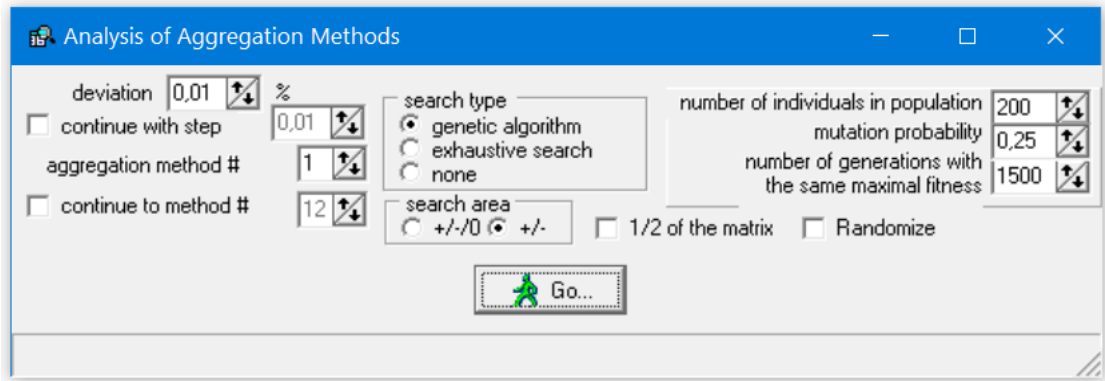

**Figure 2:** Interface of the experimental module for analyzing pairwise comparison aggregation methods

The value $\delta$ - the relative deviation of pairwise comparisons, is entered in the input field called "deviation" and can be automatically increased with a certain specified step to continuously perform a series of computations and accumulate data.

The module also provides a mode for checking the correctness of the GA results for certain specified input parameters. It is advisable to check only at small values of $n$ ($n < 6$) using a complete enumeration of the options for the values of the arguments (the elements of the PCM $A'$). For values $n \geq 6$ such checks are not advisable because they require significant computing resources.

To search for the optimum, the selection of the input parameters of the GA (number of individuals in the population, mutation probability, and number of generations with the same maximum fitness is carried out iteratively, using, if possible, a full search for verification.

In addition, the program module allows users to choose one of two modes of searching for the maximum of the function: checking the options for deviating an argument either with the option of leaving the argument unchanged (denoted as "+/-/0") or without such an option ("+/-"). Numerical studies demonstrated that it is sufficient to use a narrower range of search options when all arguments change by relative value to positive or negative values (this mode is indicated as "+/-" on the screen form shown in Fig. 2).

For experimental computations, it is also possible to use either the entire PCM or only its elements located above the main diagonal of the matrix. This mode is activated by checking the "1/2 of the matrix" checkbox.

To obtain experimental results on determining the consistency of the PCM, the existing tools for conducting an experimental study were finalized. Without changing the interface for entering the experiment parameters, a procedure for calculating the consistency index for the PCM, which serves as the GA fitness function, was added to the program module. In addition, the module is configured to search for the minimum of this fitness function, unlike the previous module, which was configured to search for the maximum.

To ensure the correctness of the experimentally obtained results, the modified program module should be used for consistency computations with the same input parameter settings as in previous computations related to the efficiency of aggregation methods.

Similarly, to the assessment of the effectiveness of aggregation, when determining the consistency of the PCMs, it was decided that it would be sufficient to further use the results of the study of the dependence of the maximum deviation of the estimates from the standard weights determined by the combinatorial method for the deviation of the elements of a perfectly consistent PCM in the range from 0 to 50% (see Fig. 3). That is, the relative errors of the expert in the pairwise comparison are simulated to be no more than 50%, which largely corresponds to the errors in real expertise.

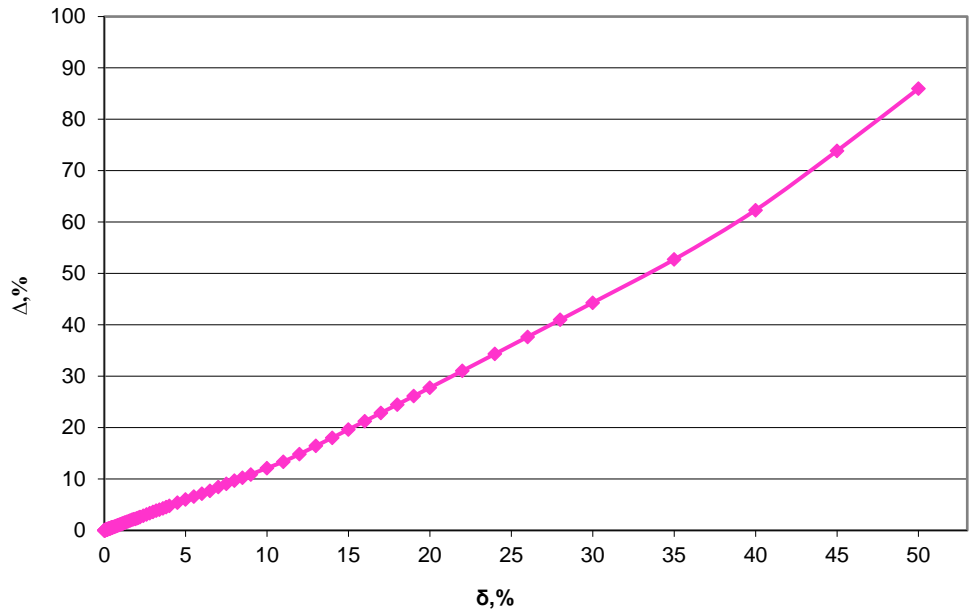

**Figure 3:** The results of the study of the dependence of the maximum deviation of estimates from the scales' standards were used

The results of the computations performed with the help of GA are shown graphically in Figure 4.

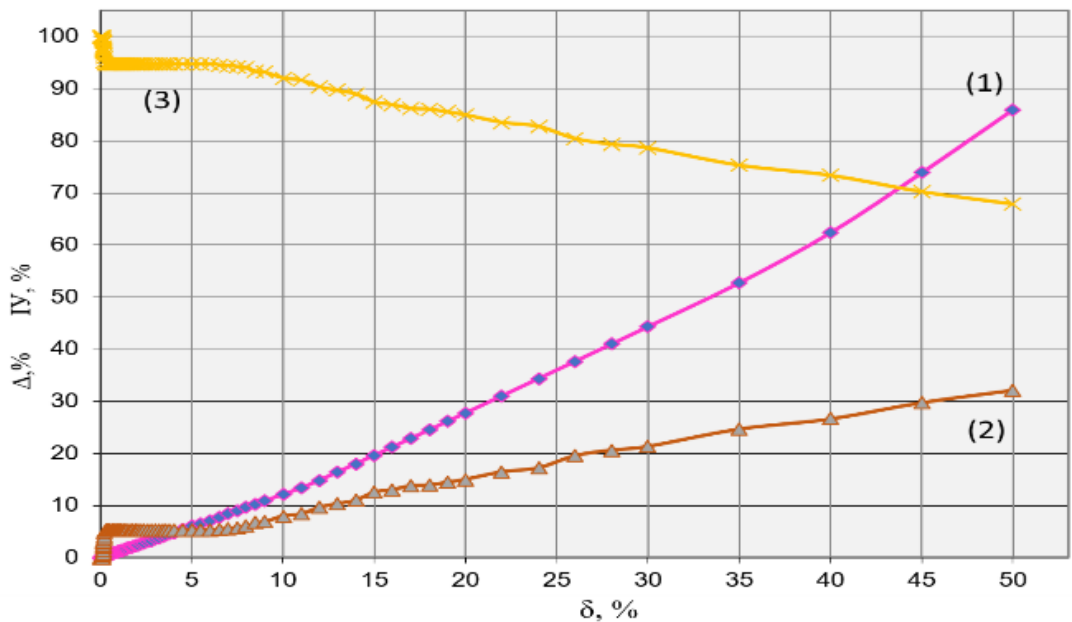

**Figure 4:** Results of determining the consistency index for the Combinatorial method of pairwise comparison aggregation

Graph (1) shows the dependence of $\Delta(\delta)$, calculated for the Combinatorial aggregation method [6, 11], (2) (dependence of the inconsistency on $\delta$ and (3)) the dependence of the consistency index on the relative estimation error $I(\delta)$. In fact, the values of the consistency index of dependence (3), shown in the figure, $I(\delta)$, match to the consistency threshold for the corresponding values of the given aggregation quality $\Delta(\delta)$, performed by the combinatorial method during the expertise. Hence, it is easy to determine the dependence $I(\Delta)$ (consistency threshold for a given $\Delta$) the permissible deviation of the resulting relative weight of the alternative from the hypothetical reference value.

It is easy to see a significant correlation between the dependencies $\Delta(\delta)$ and $(1 - I(\delta))$, i.e., as the relative error of the assessment increases, the inconsistency of expert estimates increases, which is also supported by common sense.

That is, Figure 4 shows such dependences on the elements' deviation of a perfectly consistent PCM built with reference weights (the relative error of the expert in performing pairwise comparisons):

(1) the maximum deviation from the reference values weights determined by the Combinatorial method of aggregation of PCMs;

(2) the maximum possible "inconsistency" index in percentage terms (100%–ІУ);
(3) the minimum possible consistency index in percentage terms.

The maximum possible values of (1) and (2) were computed using GA and, as noted earlier, the computed values of the dependence consistency index I(δ) represent the consistency threshold for the corresponding values of the required specified aggregation quality Δ(δ) during the expertise.

Also, it should be noted that using this approach, it is appropriate to determine the dependence of the consistency (inconsistency) indices on the requirements for the reliability of the obtained expert estimate results for various expert assessment methods for the corresponding consistency indices are used. For example, the consistency ratio for the eigenvector method used in the classical method of Analytical Hierarchy Process [12], Double Entropy Agreement Indices [2], and others.

## 5. Using the research results

The results obtained in the simulation study are best utilized when visualized within the same coordinate system as shown in Fig. 1 and Fig. 3. In fact, both dependencies $\Delta(\delta)$ and $I(\delta)$ are suitably represented with a common horizontal axis $\delta$, which is the relative error of expert pairwise comparisons (see Fig. 4).

Further, Figure 5 shows two dependencies for determining the consistency threshold according to a given Δ, which must be ensured in the expertise.

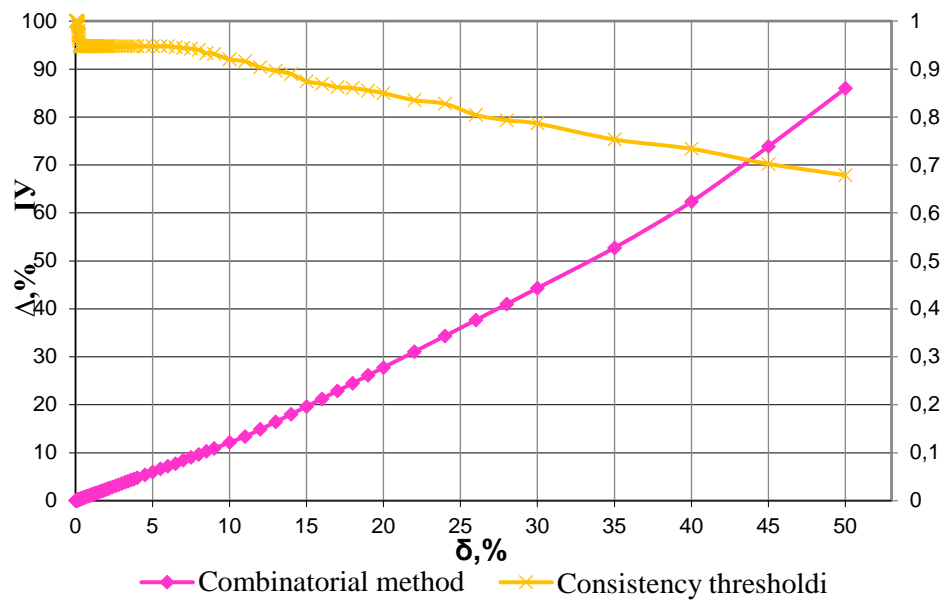

**Figure 5**: Graphs for determining the consistency threshold

Thus, using these graphs, it is possible to determine the consistency threshold at which further reasonable aggregation of expert estimates is advisable. The consistency threshold refers to the value of the consistency index below which the estimates are deemed inconsistent. In such cases, it is necessary to improve consistency before proceeding with the aggregation of these estimates.

### 5.1. A graphical way to determine the consistency threshold

Thus, the graphs of dependence $\Delta(\delta)$ and $I(\delta)$ can be employed to determine the threshold of consistency of estimates for a given value of the required level of relative deviation of the resulting aggregate values of these estimates from the reference values. Based on the graph of $\Delta(\delta)$, the value of $\delta$ is determined. It is this value of $\delta$ that is used in the dependence $I(\delta)$ to determine the value of the consistency index $I$ to be achieved, i.e., the consistency threshold. Note that the analytical determination of the consistency threshold for a given level of deviation of the resulting estimates is possible and seems appropriate in most practical applications. This method of determining the threshold will be discussed below.

### 5.2. Table values of the consistency threshold

Based on the experimentally determined dependencies $\Delta(\delta)$ and $I(\delta)$ it is not difficult to obtain the dependence $I(\Delta)$, which makes it possible to determine the consistency threshold and use it in practice. Based on the graphs shown in Fig. 5, we obtain the dependence values plotted in Fig. 6.

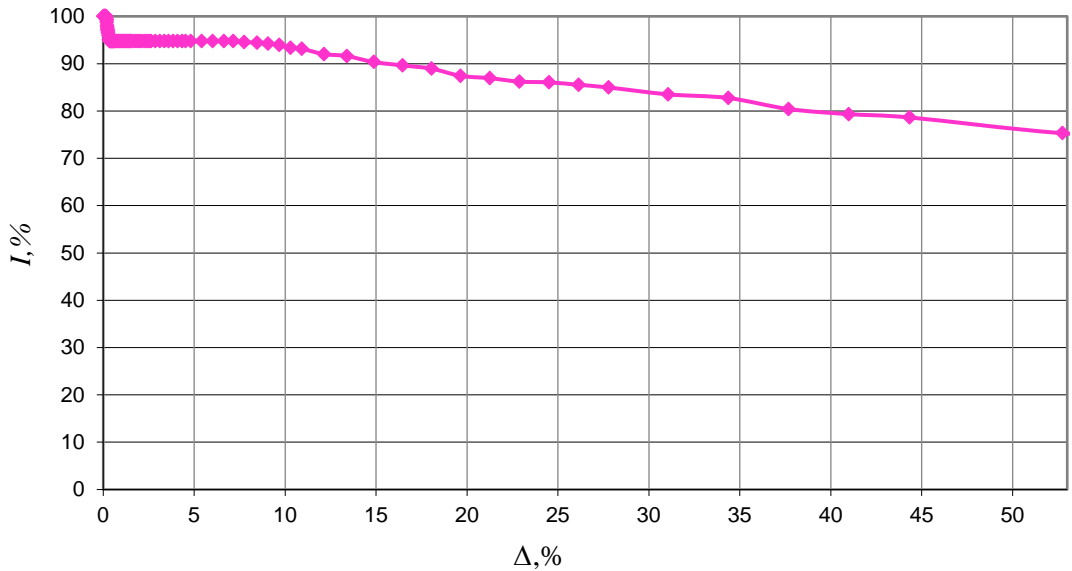

**Figure 6**: Dependence of the consistency threshold on the set Δ,%

Given a certain discretization, these dependence values can be given in a tabular form (see Table 1).

**Table 1**
**Consistency threshold depending on the required level of reliability of the expertise**

| Δ,% | < 10 | 10 | 15 | 20 | 25 | 30 | 35 | 40 | 45 | 50 |
|---|---|---|---|---|---|---|---|---|---|---|
| I | 0,95 | 0,94 | 0,9 | 0,87 | 0,86 | 0,84 | 0,82 | 0,8 | 0,78 | 0,76 |

Table 1 is useful in scenarios when the reliability of the result is an important indicator for which the requirements are set in advance.

### 5.3. Analytical expression for determining the consistency threshold

For practical applications, the analytical expression derived from the linear trend, as shown in the dependence graph in Figure 6, may also prove useful (see Figure 7).

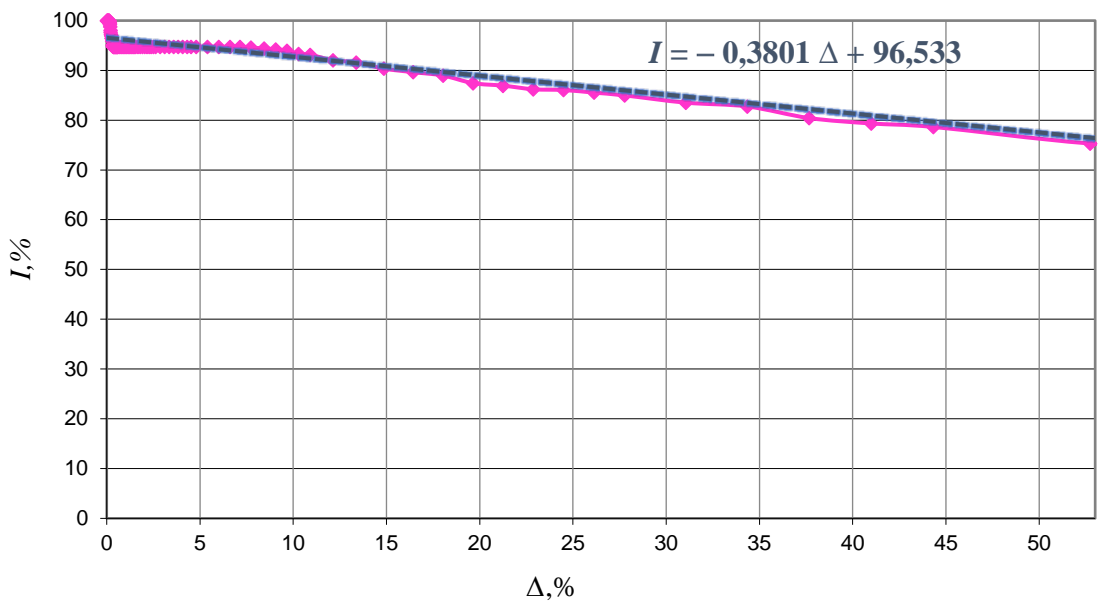

**Figure 7**: Linear trend of the consistency threshold versus Δ,%

The plot approximately follows the linear relationship $I = -0,3801\,\Delta + 96,533$ which can be used to determine the consistency threshold. The application of such experimentally calculated consistency thresholds allows increasing the reliability of the obtained estimates and improves the mechanism of conducting expertise.

# 6. Conclusions

A new approach is proposed to determine the threshold of consistency of expert pairwise comparisons sufficient to aggregate these assessments and obtain relative weights of alternatives with a given reliability. These thresholds indicate the limit of consistency at which expert estimates can be generalized (aggregated), otherwise the consistency needs to be improved. The threshold values of the consistency index were experimentally obtained depending on the required level of reliability of the resulting estimates. Determining the consistency threshold that corresponds to the specified permissible deviation of the resulting relative weight of the alternative from the hypothetical reference value allows for increasing the efficiency of the process of improving consistency and saving expensive expert resources.

# References

[1] Brunelli Matteo, A survey of inconsistency indices for pairwise comparisons, *International Journal of General Systems*, 2018, 47:8, 751-771, DOI: 10.1080/03081079.2018.1523156
[2] Olenko Andriy and Tsyganok Vitaliy, Double Entropy Inter-Rater Agreement Indices. *Applied Psychological Measurement.* 2016. Vol. 40(1). P. 37–55. DOI: 10.1177/0146621615592718
[3] Totsenko V.G. Spectral Method for Determination of Consistency of Expert Estimate Sets. Engineering Simulation. – 2000. – 17. – P.715-727.
[4] Saaty T. L. and Kearns K. P. Analytical Planning the Organization of Systems, Pergamon Press, 1985.
[5] Tsyganok V.V. Determining the effectiveness of methods of aggregation of expert estimates when using pairwise comparisons. *Data Recording, Storage & Processing.* 2009. Vol. 11, No. 2. P.83-89. (in Ukrainian)
[6] Tsyganok V.V. Investigation of the aggregation effectiveness of expert estimates obtained by the pairwise comparison method. *Mathematical and Computer Modelling.* – August 2010. – v.52, №3-4. – P.538-544. DOI: 10.1016/j.mcm.2010.03.052
[7] Tsyganok V.V., Roik P.D. Method for determining and improving the consistency of expert estimates in supporting group decision-making. *System research and information technology.* 2018., №3. - C.110-121. (in Ukrainian) https://doi.org/10.20535/SRIT.2308-8893.2018.3.10
[8] Roik P.D., Tsyganok V.V. A method for improving the consistency of expert assessments in the course of a dialog. *Data Recording, Storage & Processing.* 2018. т.20. №2. C.85-95. (in Ukrainian) https://doi.org/10.35681/1560-9189.2018.20.2.142915
[9] Tsyganok V.V. Elements of the combinatorial approach in determining the spectral coefficient of agreement of expert pairwise comparisons. *Data Recording, Storage & Processing.* - 2012. - Vol. 14, No. 2. - P.98-105. (in Ukrainian) DOI: 10.35681/1560-9189.2012.14.2.105056 https://dss-lab.org.ua/documents/Tsyganok_DRSP_2012_2(ukr).pdf
[10] Holland J.H. Adaptation in Natural and Artificial Systems. University of Michigan Press. Ann Arbor, 1975.
[11] Tsyganok V.V. Combinatorial algorithm for pairwise comparisons with expert feedback. *Data Recording, Storage & Processing.* 2000. Т.2, №2. С.92-102. (in Ukrainian) https://dss-lab.org.ua/documents/Tsyganok_DRSP-2000-2(ukr).pdf
[12] Saaty T.L., Shang J.S., An innovative orders-of-magnitude approach to AHP-based mutli-criteria decision making: Prioritizing divergent intangible humane acts. *European Journal of Operational Research*. vol. 214, issue 3. 2011. 703-715. https://doi.org/10.1016/j.ejor.2011.05.019
[13] Saaty, T. L. 1980. The Analytic Hierarchy Process. McGraw-Hill, New York.